\documentclass[aps,prl,showpacs,superscriptaddress,twocolumn]{revtex4}
\usepackage{graphics}
\usepackage[dvips]{color}
\usepackage[T1]{fontenc}
\usepackage[latin1]{inputenc}
\usepackage{graphicx}
\usepackage{amssymb}
\usepackage{rotating}
\usepackage{epsfig}

\input epsf.sty

\newcommand{\etal}{{\em et al.}}

\newcommand{\dir}{{.}}

\begin{document}

\title{Spontaneous formation of complex micelles from homogeneous solution}

\author{Xuehao He}
\affiliation{Department of Polymer Science and Engineering, Tianjin University,
300072 Tianjin, China}
\author{Friederike Schmid}
\affiliation{Fakult\"at f\"ur Physik, Universit\"at Bielefeld, D -- 33615
Bielefeld, Germany}

\begin{abstract}
We present an extensive computer simulation study of structure formation in
amphiphilic block copolymer solutions after a quench from a homogeneous state.
By using a mesoscopic field-based simulation method,
we are able to access time scales in the range of a second. 
A ``phase diagram'' of final structures is mapped out as a function of the 
concentration and solvent-philicity of the copolymers. A rich spectrum of
structures is observed, ranging from spherical and rodlike micelles and vesicles
to toroidal and net-cage micelles. The dynamical pathways leading to these
structures are analyzed in detail, and possible ways to control the structures
are discussed briefly.
\end{abstract}


\maketitle


Amphiphilic molecules in solution such as lipids or amphiphilic block 
copolymers self-assemble into a variety of structures, {\em e.g.}, 
spherical or cylindrical micelles, lamellae, and vesicles\cite{israelachvili}. 
In the case of single-component amphiphiles, vesicular and toroidal 
structures are energetically less favorable than lamellar and cylindrical 
structures due to the energy penalty for bending. Nevertheless, they may 
be stabilized by entropic or kinetic factors and still form spontaneously.
These structures provide new opportunities for designing soft materials with 
enhanced functionalities for various applications, such as complex micro 
release systems or templates for nanodevice 
fabrication\cite{shen,zipfel,zhang1,zhang2,yu1,yu2}. 
A detailed understanding of the aggregation process is crucial to understand 
and eventually control their formation.

In the past decades, a number of experimental studies have revealed the
rich diversity of micellar morphologies displayed by amphiphilic systems.
Besides spherical, rodlike or wormlike 
micelles, micelles with various special topologies have been observed, 
such as unilamellar and multilamellar vesicles, onion vesicles, 
genus vesicles\cite{shen,zipfel,zhang1,zhang2,yu1,yu2,du,he,lee},
toroidal micelles with one or several rings, and net or cage 
micelles\cite{reynhout,jiang,jain,pochan}. The experiments indicated that 
the micellar structures and size distributions not only depend on 
molecular parameters, {\em i.e.}, the chain length of the 
amphiphilic molecules, the hydrophilic-to-hydrophobic ratio, the molecular 
stiffness and the intermolecular interactions, but also on system parameters 
such as the concentration, and on kinetic factors such as the diffusion 
ability of the amphiphilic molecules and the details of the manufacturing 
process. Unfortunately, detailed dynamical information on the process of 
spontaneous micelle formation is scarce. Only a few groups have captured the
process of spontaneous vesicle formation in solutions of amphiphile 
mixtures and proposed a possible pathway of vesicle formation
\cite{egelhaaf,weiss,schmoelzer,nieh,leng}.

\begin{figure}[b]
\includegraphics[scale=0.6,angle=0]{\dir/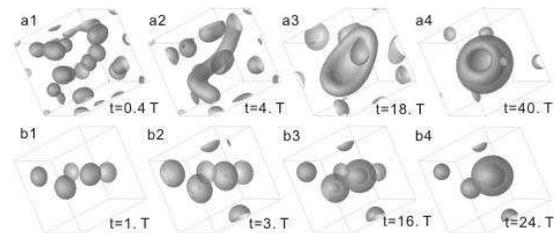}
\vspace*{-0.3cm}
\caption{\label{fig:pathways} 
Pathways of spontaneous vesicle formation in copolymer solutions.
(a) PC: Micelle coalescence (a1,a2), bilayer formation (a3), and bending (a4);
(b) PG: Micelle growth (b1,b2), internal reorganization to semivesicle (b3),
and swelling into vesicle (b4). The parameters of the simulations are 
$\chi_{BS}=0.128$ and $\Phi_p = 0.2$, (a) $\Phi_P=0.15$ (b).
Times $t$ are given in units of $T=10^4 \tau_0$.
Here and throughout the paper, structure snapshots show isodensity surfaces 
of $A$-blocks at $\Phi_A=0.625$.
}
\end{figure}

According to this ``standard'' pathway (PC), the amphiphilic molecules first
self-assemble into small spherical micelles, these then coalesce to rods, 
the rods transform themselves to bilayers, and finally, the bilayers bend 
around and close up to vesicles. The last two steps are driven by the rim 
energy of the bilayers. The mechanism has been confirmed by computer 
simulations of different coarse-grained models\cite{bernardes,noguchi,yamamoto,
devries,sevink1,uneyama}. It clearly contributes to the formation of vesicles in
amphiphilic systems. However, it cannot explain the existence of complex 
toroidal structures, since no force pushes rodlike micelles with 
two detached end caps to form rings, let alone cage structures. 
In a recent paper, we have reported the existence of an alternative pathway 
of vesicle formation in copolymer solutions\cite{he1} (PG). 
In this pathway, the micelles do not coalesce, but simply grow by attracting 
copolymers from the solution. Once a critical micelle size is exceeded, 
copolymers start to flipflop such that the micelle core becomes solvent-philic 
(``semivesicle'' state). Finally, solvent diffuses inside the core, and the 
semivesicle swells into a vesicle. The two pathways are illustrated in 
Fig.~\ref{fig:pathways} (see below for simulation details). The pathway PG
may provide a possible route to toroidal structures.

In this letter, we report on an extensive systematic study of structure
formation in a single-component amphiphilic diblock copolymer system. 
The final self-assembled structures depend strongly on the copolymer 
concentration and the interaction parameters. By varying 
the latter over a wide range, we can map out the final topologies in
a unifying ``phase diagram''. The simulations allow to investigate the 
formation process in detail. Both pathways PC and PG described above can 
be observed, depending on the copolymer concentration. We find that 
vesicles and rodlike micelles may form irrespective of the pathway, 
but toroidal structures only form {\em via} the pathway PG. Our results 
thus demonstrate that complex structure formation is not only controlled by 
the molecular packing parameters, but also, crucially, by the details of
the segregation kinetics.

Complex vesicle formation has also been studied recently by Sevink and
Zvelindovsky\cite{sevink1,sevink2}. They considered copolymers made of 
two incompatible blocks that were both basically solvent-phobic. 
As a result, the copolymers aggregated to compact, internally structured 
droplets (onion vesicles) with a relatively low solvent content 
(a few percent\cite{sevink2}). In contrast, in this work, we focus
on copolymers with strongly solvent-philic components, and on open and 
hollow structures.


We consider a system of amphiphilic diblock copolymers $P$ (copolymer volume
fraction $\Phi_P$) with solvent-phobic blocks $A$ (chain fraction $c_A$) 
and solvent-philic blocks $B$ (chain fraction $c_B = 1-c_A$), immersed in a 
solvent $S$\cite{he2}. The monomer interactions are characterized in terms 
of Flory-Huggins parameters $\chi_{AB}$, $\chi_{AS}$, and $\chi_{BS}$. 
A compressibility modulus $\kappa_H$ ensures that the local density (polymer
plus solvent) is roughly constant. The time evolution of the system is modeled 
with External Potential Dynamics\cite{maurits}, a dynamic density functional 
theory which locally conserves densities and is approximately valid for Rouse-type 
chain dynamics, but neglects hydrodynamics and reptation
(see Ref.\onlinecite{he1} for a compilation of the dynamical 
equations). The relevant dynamical model parameters are the mobility 
coefficients $D_S$ and $D_P$ of the solvent and the copolymer.

The parameter $\chi_{BS}$ and $\Phi_P$  were variable.
The other model parameters were set to $\chi_{AB} = 0.896$, 
$\chi_{AS}=1.024$, $\kappa_H=1.176$, $c_B/c_A = 0.133$, and $D_S/D_P = 17$. 
The high values of $\chi_{AS}$ and $\chi_{AB}$ ensured that the $A$-blocks 
segregate well from the $B$-blocks and the solvent. Copolymers with 
short $B$-blocks were used to stabilize bilayer structures in the regime 
where $B$ is swollen with solvent.  The remaining parameters, $D_P$ and $R_g$ 
(the unperturbed radius of gyration of the chains), set the length scale 
($r_0 := R_g/3$) and the time scale ($\tau_0:=r_0^2/D_P$) of the simulation. 
Mapping these to real copolymer solutions such as, {\em e.g.}, those studied 
in Ref.~\onlinecite{riegel} ($D_P\approx 10^{-6}$cm${}^2$s${}^{-1}$ and 
$R_g \approx 30$nm), we find that our time unit corresponds to roughly 
$\tau_0 \sim 10^{-6} s$. 

On the technical side, the parameters of the simulation were as follows:
The time step for integration of the dynamical equation was chosen
$\Delta \tau = 0.02 \tau_0$, on a spatial grid with grid size $r_0 = R_g/3$. 
The contours of the copolymer chains were discretized with $N=17$ steps. 
A small Gaussian noise 
was added to mimick the effect of thermal fluctuations\cite{he1}. 
The longest total simulation time was $7.2 \times 10^5 \tau_0$, corresponding 
to 0.72 second.

\begin{figure}[t]
\includegraphics[scale=0.8,angle=0]{\dir/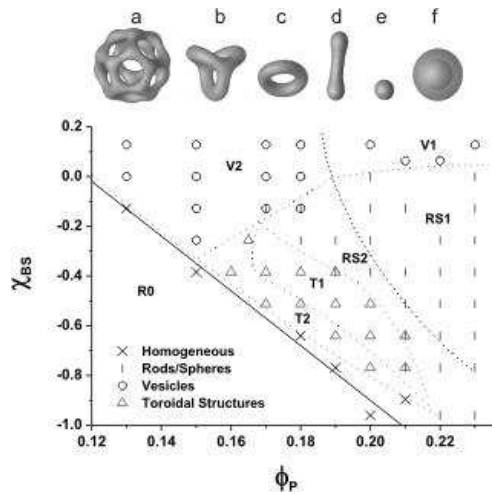}
\vspace*{-0.3cm}
\caption{\label{fig:phases} 
``Phase diagram'' of final structures after a sudden quench from an initially
homogeneous copolymer solution, for a range of solvent-philicities $\chi_{BS}$
and copolymer volume fractions $\Phi_P$. The final structures in the regions 
V1/V2 correspond to vesicles (f), RS1/RS2 to mixtures of rod and sphere 
micelles (d,e), T1 to ring micelles(c), and T2 to toroidal micelles (a,b). 
In the region R0, the solution stayed homogeneous.
The dotted lines are guides for the eye. The dashed line 
separates two dynamical regimes where the structure formation proceeds 
along different pathways: Micelle coalescence (pathway PC) in the regions 
RS1 and V1, micelle growth (pathway PG) in the regions RS2, V2, T1, and T2.
The solid line shows the function $\chi^* = 1.3 - 11. \Phi_P$ 
(see text and Fig.~\protect\ref{fig:incubation} for explanation).
}
\end{figure}


Fig.~\ref{fig:phases} shows a diagram of final structures, obtained after
quenching the system suddenly from an initially perfectly
homogeneous state. The solvent-philicity $\chi_{BS}$ ranges from 
positive to negative, in order to represent a wide class of amphiphilic
block copolymers from nonionic to ionic. In addition, the copolymer 
volume fraction $\Phi_P$ was varied in the region where interesting 
structures were observed. 

In order for structure formation to take place, the copolymer concentration
$\Phi_P$ must exceed a certain critical value, which can be identified with
the CMC (critical micelle concentration). At lower copolymer concentrations,
the translational entropy of the copolymers prevents them from aggregating.
The shape of the final structures depends on the solvent-philicity of the 
$B$-block, ($-\chi_{BS}$). At moderate $\chi_{BS}$ 
($\chi_{BS} > 0$), the system favors bilayer structures and
the copolymers aggregate to vesicles. As the solvent quality for 
the $B$-block increases ($\chi_{BS} < 0$), the $B$-block swells. Its radius
$R_B$ roughly scales with\cite{degennes} $R_B \sim (c_B N)^{3/5} (1/2 -
\chi_{BS})^{1/5}$ ($N$ is the chain length). This in turn decreases the
critical packing parameter\cite{israelachvili}, the bilayers become unstable
and give way to cylindrical and spherical structures. Consequently, the 
copolymers aggregate to rods and/or spheres in most of the parameter region. 
Close to the CMC, however, more complex structures are formed: 
Ring micelles, toroidal micelles, and even cage-like micelles.

\begin{figure}[t]
\includegraphics[scale=0.8,angle=0]{\dir/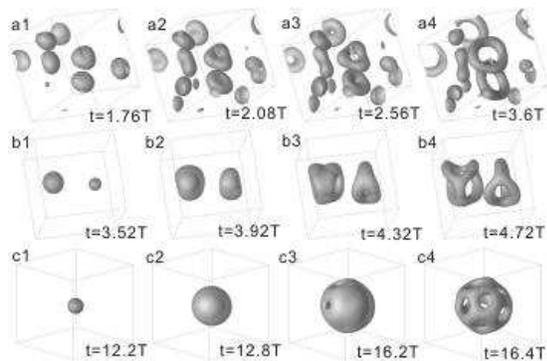}
\vspace*{-0.3cm}
\caption{\label{fig:toroidal} 
Formation of toroidal structures at
$\chi_{BS} = -0.512$:
(a) Rings ($\Phi_P = 0.2$),
(b) Toroidal micelles ($\Phi_P = 0.18$),
(c) Cage micelles ($\Phi_P = 0.17$).
Times $t$ are given in units of $T=10^4 \tau_0$.
}
\end{figure}

To understand why these complex micelles appear, one must inspect the
pathways of structure formation in more detail. Both pathways to vesicle 
formation, PC and PG, are observed in our system (Fig.~\ref{fig:pathways}).
Likewise, rod formation also proceeds {\em via} the two distinct pathways
micelle coalescence (PC), or (anisotropic) micelle growth (PG), depending
on the copolymer volume fraction (Fig.~\ref{fig:phases}): Coalescence takes
place at $\Phi_P \stackrel{>}{~}0.2$. Rings and toroidal micelles emerge at 
much lower copolymer volume fraction, and their formation is clearly driven by 
a growth mechanism. Fig.~\ref{fig:toroidal}a shows the pathway to ring
formation, where spherical micelles first grow into small disks, a hole
then nucleates at the center of the disks, and finally, the perforated
disks evolve into rings. Even closer to the CMC, the same mechanism leads
to the formation of toroidal micelles: The micelles first grow into
semivesicles or small vesicles, then several holes appear in the vesicle
shells, until finally, the perforated vesicles grow into toroidal micelles
(Fig~\ref{fig:toroidal}b,c). The number of holes depends on the size of the
embryo vesicle at the time of breakup. Close to the CMC, the dynamical
stability of large vesicles increases: The initial number of micellar nuclei 
is small and they are far apart, hence the vesicles are not perturbed by the 
environment and break up late. As a result, large cage micelles can be 
obtained in the vicinity of the CMC (Fig.~\ref{fig:toroidal}c).

\begin{figure}[t]
\includegraphics[scale=0.5,angle=0]{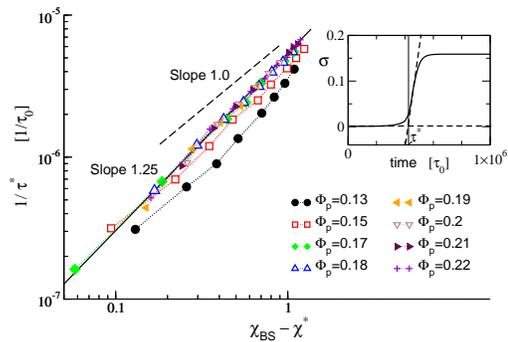}
\vspace*{-0.3cm}
\caption{\label{fig:incubation} 
Inverse incubation time $1/\tau^*$ as a function of $\chi_{BS}$ for different
copolymer volume fractions $\chi^*$ as indicated. The values $\chi_{BS}$ have
been shifted by $\chi^* = 1.3-11 \Phi_P$. The solid line corresponds to the 
function $y=5.3 \cdot 10^{-6} x^{1.25}$, and the dashed line shows $y \propto x^{-1}$ 
for comparison. Inset: Evolution of the order parameter $\sigma$ with time at 
$\chi_{BS}=0$, $\Phi_P=0.17$. The incubation time is defined as the time
where the tangent of $\sigma(t)$ at the inflection point and the base 
line $\sigma \equiv 0$ intersect each other.
}
\end{figure}

Next we examine the early stages of micelle aggregation. To this end, we 
define the segregation parameter 
$\sigma = \int {\rm d}r |\Phi_A(r) + \Phi_B(r) - \Phi_P|/V$,
where $\Phi_A(r)$ and $\Phi_B(r)$ are local monomer densities, and
$V$ is the volume. Looking at $\sigma$ as a function of time, we find that
the copolymer segregation proceeds in an almost step-like fashion after 
a well-defined ``incubation time'' $\tau^*$ (Fig.~\ref{fig:incubation}, inset). 
Fig.~\ref{fig:incubation} shows the incubation times as a function of
$\chi_{BS}$ for different copolymer volume fractions $\Phi_P$. Remarkably,
most curves collapse onto a single power law function
$1/\tau^* = A \cdot (\chi_{BS} - \chi^*)^\alpha$ with the exponent
$\alpha \approx 1.25$ after a simple shift of 
$\chi_{BS}$ by $\chi^*(\Phi_P) = 1.3-11 \Phi_P$. 
Only for the lowest copolymer volume fractions, $\Phi_P=0.13$ and $\Phi_P=0.15$, 
do the data fail to collapse; however, the slopes of these curves, 
shifted by $\chi^*$ and plotted in a double logarithmic way, are 
still comparable to $\alpha$.

A similar power law behavior has been observed previously in a simulation
study of vesicle formation in two dimensions\cite{he1}. It was explained
in terms of the Cahn-Hilliard theory for spinodal decomposition, and 
$\chi^*$ was identified with the spinodal for macrophase separation
between polymer and solvent. In the present case, however, $\chi^*(\Phi_P)$ 
is significantly lower than that spinodal\cite{note1}, it seems 
rather related to the CMC (cf. Fig.~\ref{fig:phases}). It is
worth noting that data collapse and power law behavior is observed in a 
range of $\Phi_P$ {\em regardless} of the final structure and the dynamical 
pathway of structure formation. This suggests that the characteristics
of the initial stage of segregation are universal and related to 
a spinodal-type instability. 

Once created, the micelles grow by attracting copolymers from the solution. 
In the pathway PC, the growth is supplemented by micelle coalescence. The 
question arises under which conditions this happens. In fact, most fusion 
events take place at early times (see Fig.~\ref{fig:pathways} a1,a2). 
At later stages, they are impeded by two factors: 
The formation of a well-segregated, swollen $B$-corona at the surface of 
the micelles, and the emerging copolymer depletion zone around the micelles. 
Hence the density of micelle nuclei at the early stage is a candidate
quantity that might select between pathways. The crossover between pathways
is observed at the copolymer volume fraction $\Phi_P \approx 0.2$.
Taking into account that a large fraction of copolymers assembles into droplets 
almost simultaneously (at the time $\tau^*$), and that these droplet nuclei 
roughly have the diameter $2 R_g$, we can estimate the average distance 
$D$ between droplets at given copolymer content\cite{note2} {\em via} 
$(2R_g/D)^3 \sim \Phi_P$. At $\Phi_P = 0.2$, $D$ is of the order $R_g$. 
Hence the copolymers in solution are in contact with several droplets,
they are attracted by all of them, and can serve as bridges that mediate 
fusion. 
 
In the last stage, the structures ripen.
As long as they are still small (semivesicle state), classical 
Oswald coarsening is observed (Fig.~\ref{fig:pathways}a3-a4, b3-b4):
Small structures dissolve, large structures grow, driven by the competition 
of bulk and surface free energy. Once the structures have locally assumed 
their favored toroidal or bilayer structure, the Oswald process stops
and the ripening is governed by much weaker thermodynamic
forces, such as, {\em e.g.}, those associated with the bending energy. 
The time scales of these processes are very slow and out of reach for our
simulation method. Therefore, we have carried out a set of simulations 
using external potential dynamics with locally non-conserved (but globally 
conserved) densities. This dynamical model is less realistic, but much faster, 
such that we could also assess later stages of the aging process. 
Specifically, we studied the evolution of a system containing two vesicles 
with different initial sizes in two and in three dimensions. 
In the case of ring micelles or two dimensional vesicles, the contribution
of the bending energy favors a uniform size distribution: The energy of a 
single ring of radius $R$ is proportional to $k/R$ ($k$ being the bending 
rigidity), and the total energy for a fixed number of rings is minimal if
all rings have the same diameter. Indeed, our two dimensional simulations 
showed that the sizes of the two rings converged in the course of the
simulation: The small ring grew at the expense of the large ring. 
For three dimensional vesicles, the situation is different:
The bending energy of a vesicle is $4 \pi k$, independent of its size. 
The total bending energy only depends on the number of vesicles, 
not on their size distribution. Consequently, we did not observe any sign 
of size uniformization, nor size disproportionation, in the three dimensional
simulations. The same behavior has been observed in experiments\cite{cheng}.


To summarize, we have investigated the formation of toroidal micelles in 
copolymer solutions, and shown that such micelles may form in the vicinity 
of the CMC by a pathway that proceeds {\em via} the nucleation, growth, 
and subsequent breakup of vesicles. The competition of copolymer aggregation 
and self-assembly generates a kinetic trap that opens a route to manufacturing 
highly complex metastable structures. Several methods to control these 
kinetic traps are conceivable: Controlling the number of initial nuclei by 
planting seeds (prenucleation)\cite{fraaije, he3}, quenching from different 
initial states({\em e.g.}, quenching from a vesicle state), doping with 
additional components to tune the bilayer properties, or working with 
amphiphilic mixtures.

X.H. thanks the Alexander von Humboldt foundation for a research fellowship. 
The simulations were carried out at the Paderborn center for parallel
computing.


\begin{thebibliography}{66}
\bibitem{note1}
The spinodal line is given by\protect\cite{he1}
$\chi_{BS}^{\mbox{\tiny sp}} = 
\chi_{AB} c_A - \chi_{AS} c_A/c_B + 1/(2 c_B (1-\Phi_P))
+ 1/(2 c_B N \Phi_P)$.
\end{thebibliography}
\end{document}